\documentclass{aa}

\usepackage{graphicx}
\usepackage{amsmath} 
\usepackage{amssymb}
\usepackage{lscape}
\usepackage{natbib}
\usepackage{pifont}

\bibpunct{(}{)}{;}{a}{}{,}

\usepackage{txfonts}

\begin{document}

   \title{A deep optical/near-infrared catalog of Serpens}

  \author{L. Spezzi\inst{1} 
   \and B. Mer\'in\inst{2}
   \and I. Oliveira\inst{3,4}
   \and E.F. van Dishoeck \inst{4,5}
   \and J.M. Brown \inst{5}
   }

 \offprints{L. Spezzi, \email{lspezzi@rssd.esa.int}}

\institute{Research and Scientific Support Department, European Space Agency (ESA-ESTEC), P.O. Box 299, 2200 AG Noordwijk, The Netherlands
\and European Space Astronomy Center, European Space Agency (ESA-ESAC), P.O. Box Apdo. de correos 78, 28691 Villanueva de la Ca\~nada, Madrid, Spain
\and California Institute of Technology, Division for Geological and Planetary Sciences, MS 150-21, Pasadena, CA 91125, USA
\and Sterrewacht Leiden, Leiden University, P.O. Box 9513, 2300 RA Leiden, The Netherlands
\and Max Planck Institut f\"ur Extraterrestrische Physik, Giessenbachstrasse 1, 85748 Garching, Germany
}

   \date{Received ; accepted }

\abstract
% context heading (optional), leave it empty if necessary 
{ }
{We present a deep optical/near-infrared imaging survey of the Serpens molecular cloud. 
This survey constitutes the complementary optical data to the Spitzer 
"Core To Disk" (c2d) Legacy survey in this cloud.}
{The survey was conducted using the Wide Field Camera at the Isaac Newton Telescope. 
About 0.96~square degrees were imaged in the $R$ and $Z$ filters, 
covering the entire region where most of the young stellar objects identified by 
the c2d survey are located. 26524 point-like sources were detected in both 
$R$ and $Z$ bands down to $R\approx$24.5~mag and $Z\approx$23~mag with a signal-to-noise ratio better than 3. 
The 95\% completeness limit of our catalog corresponds to 0.04~M$_{\odot}$ 
for members of the Serpens star forming region (age 2~Myr and distance 260~pc) 
in the absence of extinction. Adopting the typical extinction of the observed area 
(A$_V \approx$7~mag), we estimate a 95\% completeness level down to M$\approx$0.1~M$_\odot$.
The astrometric accuracy of our catalog is 0.4~arcsec with respect to the 2MASS catalog.}
{Our final catalog contains J2000 celestial coordinates, magnitudes in the $R$ and $Z$ bands 
calibrated to the SDSS photometric system and, where possible, $JHK_S$ magnitudes from 2MASS 
for sources in 0.96~square degrees in the direction of Serpens. 
This data product has been already used within the frame of the {\bf c2d Spitzer Legacy Project} 
analysis in Serpens to study the star/disk formation and evolution in this cloud; 
here we use it to obtain new indications of the disk-less population in Serpens.}
{ }


\maketitle

\section{Introduction \label{intro}}

The Spitzer Legacy Survey ``Molecular Cores to Planet Forming Disks'' \citep[c2d;][]{Eva03} 
offered a singular opportunity for a major advance in the study of star and planet formation. 
Thanks to its sensitivity and vawelength coverage, Spitzer allowed us to address for the first time 
long-standing challenges such as disk formation and dispersal, the physical 
and chemical evolution of the circumstellar material and, in particular, 
to probe the inner planet-forming region of disks on the basis 
of statistically significant samples \citep[see, e.g.,][]{Lad06}.

One of the star-forming regions included in the c2d survey is the Serpens molecular cloud. 
Because of its proximity \citep[260~pc;][]{Str96} and young age \citep[2-6 Myr;][]{Oli09}, 
this cloud is particularly well-suited for studies of very young low-mass stars and 
sub-stellar objects. The c2d survey in Serpens has provided evidence of 
sequential star formation in this cloud progressing from SW to NE and 
culminating in the main Serpens Core with its cluster of
Class 0 objects \citep{Kaa04,Har07}; moreover, {\bf the surface density} of young 
stars in this region is much higher, by a factor of 10-100, than that 
of the other star-forming regions mapped by the c2d team \citep{Eva09} 
and includes {\bf 22\%} of the c2d sources classified as ``transitional'' disks. 
This makes Serpens the best region for obtaining a complete, well-defined sample of multi-wavelength 
observations of young stars and sub-stellar objects in a possible evolutionary 
sequence to build up a ``template'' sample for the study of disk evolution 
up to a few Myr within a single, small, well defined region. 
To this aim, the c2d Team has conducted several surveys, from X-ray to millimeter wavelengths, 
and spectroscopic follow-ups of the newly discovered population of young stars in Serpens, 
making this cloud only the third star-forming region after Taurus and IC~348 for which such 
an unbiased dataset exists \citep{Goo04,Har07,Eno07,Oli09,Oli10}. 

In this paper we present the optical/near-infrared (NIR) imaging data collected 
within the frame of the c2d survey in Serpens.
These data represent one of the critical ingredients to anchor the studies of envelopes and disks 
to the properties of the central stars, including those without infrared excess.
Indeed, deep XMM-Newton data of the same field revealed a sample of 
new sources, mostly candidate weak-line T Tauri stars (WTTSs), 
half of which have no counterpart in other catalogs \citep{Bro10}. 
Moreover, \citet{Com09} recently reported on a large-scale optical 
survey of the Lupus star-forming complex which, in combination with
NIR data from the 2MASS catalog, unveiled a large population of stars 
and brown dwarfs (BDs) that have lost their inner disks on a timescale of a few Myrs or less. 
This discovery stresses the important unknowns that still persist
in the observational characterisation of young very low-mass objects 
{\bf and in the time-scales and mechanisms for disk dissipation}.

The outline of this paper is as follows: in Sect.~\ref{sec_obs} and \ref{sec_red} 
we describe the observations and data reduction procedure, with particular emphasis on the 
photometric completeness and astrometric accuracy of the final optical/NIR catalog.
In Sect.~\ref{use} we discuss the use of this catalog within the frame 
of the c2d Team's analysis in Serpens and investigate its disk-less population. 
Our conclusions are drawn in Sect.~\ref{conclu}.

\section{Observations \label{sec_obs}}

The imaging observations presented in this work were carried out from 13 to 17 May 2008 
using the Wide Field Camera (WFC) at the 2.5m Isaac Newton Telescope (INT), which is 
located at the Roque de Los Muchachos Observatory (La Palma, Spain). 
The data were collected by Ignas Snellen as part of an observing practicum for 
Leiden University astronomy students.

The WFC is a four-chip mosaic of thinned AR coated EEV 4K$\times$2K devices.
Each CCD has an useful imaging area of 2048$\times$4100 pixels with a pixel scale of 
0.333~arcsec/pixel, covering a total field of view 34$\times$34~arcmin 
with small gaps of $\sim 20''$ between adjacent chips. The average CCD read-out noise and 
gain are 7~e$^-$ and 2.8~e$^-$/ADU, respectively. 

\begin{figure*} 
\centering
\includegraphics[width=16cm,height=14cm]{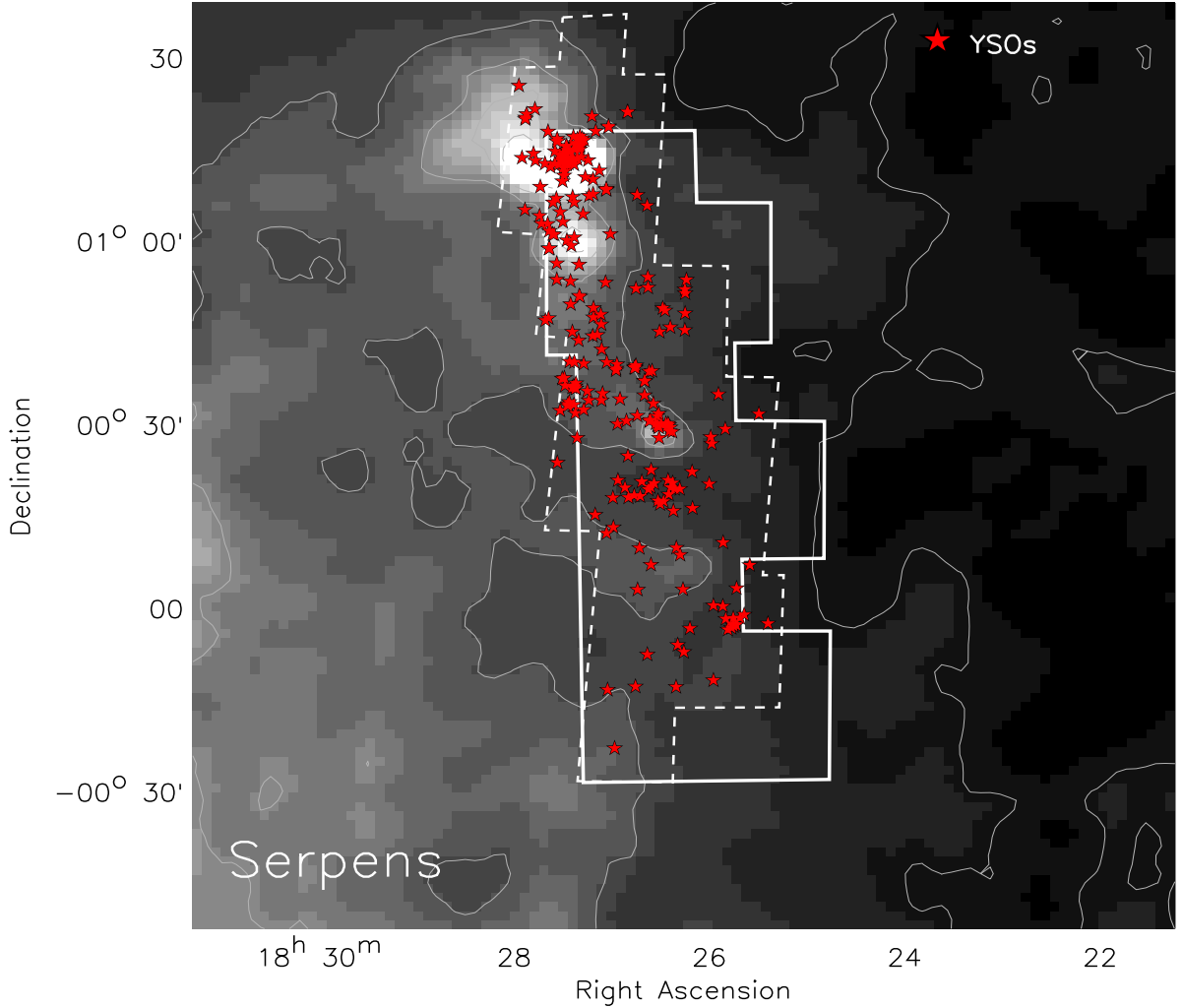}
\caption{IRAS 100$\mu$m dust emission map of the Serpens dark cloud. 
The contours, from 55 to 190 MJy $\times$ sr$^{-1}$ in steps of 
9 MJy $\times$ sr$^{-1}$, are also drawn (white lines). The continuous line 
defines the area covered by the three INT+WFC pointings, while the dashed line 
is the area observed by Spitzer {\bf as part of the c2d Survey}. 
The stars are the YSO candidates identified by the Spitzer c2d Survey.}
\label{fig_obs}
\end{figure*}

A total of 3 fields were observed in the direction of Serpens 
(R.A.=18$^h$29$^m$49$^s$, Dec.=+01$^d$14$^m$48$^s$), 
covering the entire region where most of the young stellar objects (YSOs) identified by 
the Spitzer c2d survey are located (Fig.~\ref{fig_obs}). 
Images were obtained through the $R$ ($\lambda_c$=6260~\AA, 
FWHM=1380~\AA) and $Z$ ($\lambda_c$=9100~\AA, FWHM=1370~\AA) 
filters approximately covering the Sloan Digital Sky Survey (SDSS) $r'$ and $z'$ bands \citep{Fuk96}.
The total sky-area observed in each filter is 0.96~square degrees. 
Each observation in each filter was split into several individual exposures (ditherings), 
shifting the telescope pointing by $\sim$1~arcmin 
between consecutive exposures; this allows us to cover the gaps between 
the CCDs and avoid saturation of bright sources in the field. 
The exposure time for each dithering was 120~sec in both the $R$ and $Z$ bands. 
In order to recover the photometry for bright sources saturated 
in these long-time exposures, a series of short-time exposures 
(i.e. 10~sec per dithering in both filters) of the same sky-area was also performed. 
The summary of these observations is reported in Table~\ref{tab:obs}. 
A standard Landolt field was also observed in both filters 
for absolute flux calibration purposes (Sect.~\ref{phot_cal}).

%---------------------------------- RA.vs.DEC-------------------
%-------------------------------------------------------------- 

\begin{table}
\caption[ ]{\label{tab:obs} Journal of the observations.}
\begin{center}
\scriptsize
\begin{tabular}{cccccc}  
\hline
Field             & Date       &  Filter  & T$_{\rm exp}$ & Seeing & Air Mass \\
(R.A.,Dec.)       & (d/m/y)    &          & (min)         & (")    &          \\
\noalign{\medskip}
\hline
\noalign{\medskip}
Serpens\_1             & 13/05/08 & R     & 120$\times$4 & 1.5 & 1.50 \\  
(18:28:43,+00:24:46)   & 17/05/08 & R     &  10$\times$1 & 2.0 & 1.14 \\  
                       & 15/05/08 & Z     & 120$\times$9 & 1.5 & 1.17 \\  
                       & 17/05/08 & Z     &  10$\times$1 & 2.1 & 1.14 \\  
\hline
Serpens\_2             & 14/05/08 & R	  & 120$\times$9 & 1.6 & 1.15 \\
(18:28:57,+01:02:54)   & 17/05/08 & R	  &  10$\times$1 & 2.0 & 1.15 \\
                       & 15/05/08 & Z     & 120$\times$4 & 1.5 & 1.16 \\  
                       & 17/05/08 & Z     &  10$\times$1 & 1.5 & 1.15 \\  
\hline
Serpens\_3             & 14/05/08 & R	  & 120$\times$9 & 1.6 & 1.15 \\
(18:28:36,-00:09:51)   & 17/05/08 & R	  &  10$\times$1 & 1.6 & 1.15 \\
                       & 17/05/08 & Z     &  10$\times$1 & 2.3 & 1.15 \\  
\hline
L~104$^\dag$           & 13/05/08 & R     &   2$\times$1 & 1.2 & 1.22 \\
(12:41:68,-00:34:11)   & 13/05/08 & Z     &   5$\times$1 & 1.3 & 1.19 \\
\noalign{\medskip}
\hline
\end{tabular}
\end{center}
$^\dagger$ \footnotesize{{\bf Standard star field.}}
\end{table}

\section{Data reduction \label{sec_red}}

The raw images were processed using the IRAF\footnote{IRAF is 
distributed by NOAO, which is operated by the Association 
of Universities for Research in Astronomy, Inc., under 
contract to the National Science Foundation.} \emph{mscred} package 
and a number of scripts \emph{ad hoc} developed both under 
IRAF and under IDL\footnote{Interactive Data Language.}. 
We followed the steps for the WFC data reduction pipeline \citep{Irw01} 
developed by the Cambridge Astronomical Survey Unit (CASU), 
responsable for the processing and archiving of the dataset 
from the INT Wide Field Survey \citep[WFS;][]{McM01}.

\subsection{Pre-reduction \label{prered}}

Since the bias level tends to vary, the images were first corrected 
for overscan and trimmed using the bias and trim sections as specified in the FITS headers. 
For each night in which observations for our program were performed, 
twilight flat frames were then combined to obtain the  
night master flat, which was then used to correct the science images. 
The corrected images are essentially linear to $\sim$2\% over the full range. 
Bad pixels and partial columns have been replaced using 
the bad pixel mask files available from the WFS 
homepage\footnote{http://www.ast.cam.ac.uk/~wfcsur/technical/pipeline/}.

The $Z$-band images suffer from significant sky fringes. 
In order to remove them, we subtracted from each $Z$-band image 
the fringing pattern frame available from the WFS homepage 
scaled by a specific factor to account
for the amplitude of the fringes in the individual science frames.

\subsection{Astrometry and co-addition of images \label{astrom}}

The astrometric calibration and relative flux scaling between ditherings 
were obtained using the c-version of ASTROMETRIX\footnote{See also: 
\emph{http://www.na.astro.it/$\sim$radovich/wifix.htm}} (M. Radovich, private communication). 
This tool performs a global astrometric solution that takes overlapping
sources falling on adjacent CCDs in different ditherings into account. 
For each pointing, the astrometric 
solution was computed using the USNO-B1.0 
catalogue \citep{Mon03} as a reference. Within the global astrometry process, the astrometric 
solution was constrained for each CCD by both the positions from the 
USNO-B1.0 catalogue and those from overlapping sources in all the other CCDs. 

\begin{figure} 
\centering
\includegraphics[width=9cm,height=9cm]{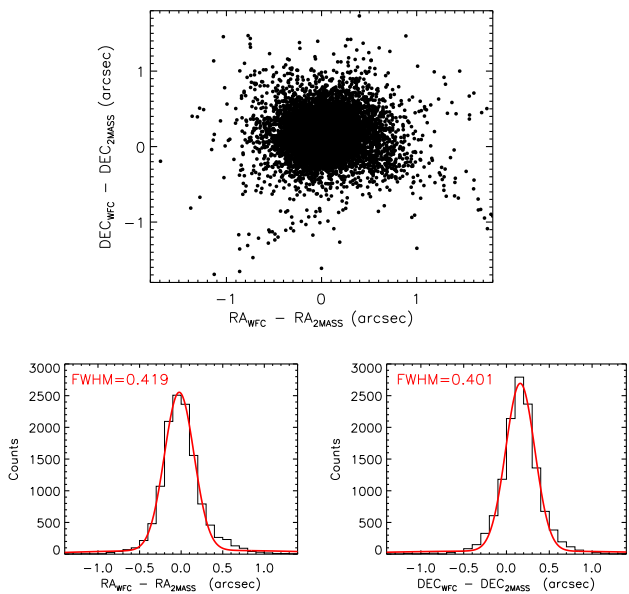}
\caption{Residuals of the coordinates obtained by us for the sources in the 
surveyed area in Serpens with respect to those from the 2MASS catalog.}
\label{match_2M}
\end{figure}

The co-addition of the dithered images for a given filter and 
pointing was performed using the SWARP tool \citep{Ber08}. 
The final stacked image is a $6k \times 6k$ frame 
where each pixel value is the median flux of the 
co-added ditherings normalised to the total exposure 
time and relative to the airmass and atmospheric 
transparency of the first frame in the dithering set. 
The absolute astrometric precision of our images is about 0.4~arcsec, 
slightly lower than RMS accuracy of the USNO-B1.0 catalogue (0.2~arcsec); 
the astrometric precision has been also confirmed by a cross-check 
with the 2MASS point-source catalog (Fig.~\ref{match_2M}). 
The internal RMS, computed from overlapping sources in different exposures, 
is within 0.05 arcsec, indicating the good performance of ASTROMETRIX.

%---------------------------------- match 2MASS-------------------
%-------------------------------------------------------------- 

\subsection{Photometric calibration\label{phot_cal}}

Instrumental magnitudes were reported to the standard SDSS photometric system \citep{Fuk96}. 
To this aim, the {\bf Stetson's standard star field L~104 \citep{Ste00}} was observed in the $RZ$ filters.
By using the IRAF package \emph{photcal}, 
we first performed the aperture photometry for the standard stars, 
obtaining their instrumental magnitudes ($r_0$ and $z_0$) 
corrected for atmospheric extinction and normalised to the exposure time.
Then, the transformation coefficients, namely zero point ($ZP$) 
and colour term ($c$), from the WFC-INT system to the SDSS standard 
system were determined by a linear fitting of the following equations:
\begin{eqnarray}
r'=r_0 + c_R \cdot (r_0-z_0)+ZP_R  \\
z'=z_0 + c_Z \cdot (r_0-z_0)+ZP_Z 
\label{cal_RZ}
\end{eqnarray}
where $r'$ and $z'$ are the standard magnitudes of Landolt's stars in the SDSS system.

The mean transformation coefficients determined in our observing run 
are reported in Table~\ref{tab:coeff}. 

Using these coefficients, instrumental magnitudes (see Sect.~\ref{catalog}) 
for the observed sources in Serpens have been converted to the SDSS photometric system, 
which is nearly an AB system. Thus, magnitudes in our catalog can be turned 
into flux densities using the correction from SDSS zeropoints to AB zeropoints and the AB zeropoint flux 
density\footnote{See \emph{http://www.sdss.org/dr6/algorithms/fluxcal.html}.}: 

\vspace{0.3cm}

correction $ZP_R$ = 0 mag

$F^0 _{r'}$ = 2.7769E-12 \rm{$W \cdot cm^{-2} \cdot \mu m^{-1}$}

\vspace{0.3cm}

correction $ZP_Z$ = 0.02 mag

$F^0 _{z'}$ = 1.3153E-12 \rm{$W \cdot cm^{-2} \cdot \mu m^{-1}$}

% --------------------------------------------- Table ----------------------------
\begin{table}
\caption[ ]{\label{tab:coeff} Mean photometric calibration coefficients for our WFC-INT observing run.}
\begin{center}
\small
\begin{tabular}{cccc}  
\hline
Filter & {\it K}$^\dagger$ & {\it ZP} & {\it c} \\
\noalign{\medskip}
\hline
\noalign{\medskip}
$R$   &  0.0734 & 25.30$\pm$0.03 & -0.033$\pm$0.024  \\
$Z$   &  0.0103 & 23.52$\pm$0.03 &  0.021$\pm$0.022  \\
\noalign{\medskip}
\hline
\end{tabular}
\end{center}
$^\dagger$ \footnotesize{{\bf {\it K} = mean atmospheric extinction coefficients for La Palma.}}\\
$^\ddagger$ \footnotesize{{\bf{\it ZP} = photometric zero point (see Eq.~\ref{cal_RZ}).}}\\
$^*$ \footnotesize{{\bf{\it c} = photometric colour term (see Eq.~\ref{cal_RZ}).}}
\end{table}
%-------------------------------------------------------------------------------------

\subsection{The catalog extraction \label{catalog}}

The source extraction and photometry from 
each stacked image in each filter were performed 
by using the 2.5 version of the SExtractor tool by \citet{Ber96}. 
SExtractor exploits the aperture photometry technique, 
which is the faster and best approach for uncrowded fields such as 
the one considered here. In particular, we adopted the 
SExtractor adaptive aperture magnitudes ($mag_{auto}$), 
which are estimated from a flexible elliptical aperture 
around each detected object; this method is expected to give the 
most precise magnitudes for very faint objects which may 
appear extended. The detection threshold was set 
in order to select all the sources having a 
signal-to-noise ratio ($\sigma$) greater than 3. 
The background is locally estimated from a ring shaped 
region surrounding the star.

The output catalogs contain for each source an identification number, 
right ascension and declination in degree at J2000, instrumental 
magnitudes and relative errors, and two SExtractor 
morphological parameters, namely the extraction flag (FLAGS) 
and the isophotal area above the threshold (ISOAREA\_IMAGE).
These last parameters were used to clean the catalogs from 
spurious detections, such as cosmic ray hits, bad pixels, 
saturated or truncated sources too close to the image 
boundaries, etc.

\begin{figure} 
\centering
\includegraphics[width=10cm,height=8cm]{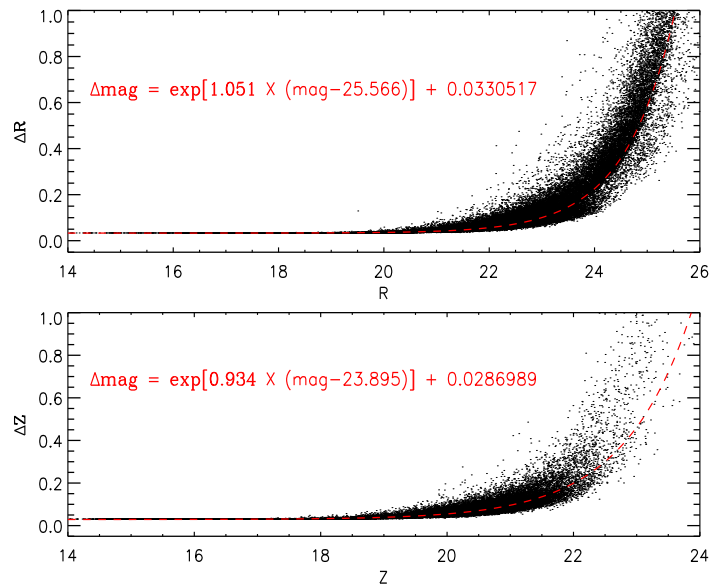}
\caption{Photometric errors versus magnitudes and relative 
exponential fits for all the point-like sources detected in the 0.96~square 
degrees area surveyed in Serpens. Photometry is from our long-time exposures (Sect.\ref{sec_obs}).}
\label{fig:err}
\end{figure}

In Figure~\ref{fig:err} we show as an example the internal photometric errors 
of all the point-like sources detected in our long-time exposures plotted 
against the magnitude for the $RZ$ filters; 
the relative exponential fits are over-plotted. 
Table~\ref{tab:mag_limits} summarises the saturation limit and 
the limiting magnitude achieved at the 10$\sigma$, 5$\sigma$ and 3$\sigma$ level 
in each filter for both the long time exposures and the short ones. 

Our final catalog contains celestial coordinates (R.A. and Dec. at J2000) 
and $RZ$ photometry for each point-like object detected in our survey 
above the 3$\sigma$ level. Each entry in the catalog was further complemented with
near-infrared $JHK_S$ photometry by cross-matching it with the
2MASS Point Source Catalog \citep[limiting magnitude $K_S \approx$15.5;][]{Skr06}. 
A matching radius of 2~arcsec was defined, which appears to be sufficient
given that more than 90\% of the matches corresponded to
differences between the 2MASS positions and those determined
by us of less than 1~arcsec. When more than one 2MASS source was
found within the 2~arcsec circle, the one closer to the position derived
from our observations was chosen as the counterpart.

%---------------------------------- errors-------------------
%-------------------------------------------------------------- 

%---------------------------------- Table mag limits-------------------
\begin{table*}
\caption[ ]{\label{tab:mag_limits} Number of stellar sources (N$_S$) 
detected in both filters above the 3$\sigma$ level and limiting magnitudes in each filter 
at 10$\sigma$, 5$\sigma$, 3$\sigma$ and 95\% completeness level.}
\begin{center}
\begin{tabular}{cccccccc}
\hline
   & N$_S$ & Filter & Mag Sat. &  Mag 10$\sigma$ &  Mag 5$\sigma$ &  Mag 3$\sigma$ & Mag (C=95\%) \\
\noalign{\medskip} \hline
Long Time Exposures  & 26524 & $R$ & 16.00 & 22.90 & 23.87 & 24.49  & 22.30 \\
                     &       & $Z$ & 14.50 & 21.12 & 22.11 & 22.73  & 19.20 \\
   			        	              	   		     																	     
\noalign{\medskip} \hline
Short Time Exposures & 3709  & $R$ & 12.50 & 19.82 & 20.84  & 21.50 & 18.30 \\
                     &       & $Z$ & 10.20 & 18.61 & 19.41  & 19.91 & 16.10 \\

\noalign{\medskip} \hline
\end{tabular}
\end{center}
\end{table*}
%------------------------------------------------------

\subsection{Completeness \label{psf_compl}}

The completeness of our catalogue was estimated in the standard way by inserting artificial stars into 
the images and recovering them using the same extraction procedure as for the real objects (Sect.~\ref{catalog}); 
the fraction of recovered artificial objects provides a measure of the completeness. 

We used the DAOPHOT~II standalone package to perform the exercise \citep{Ste87}. 
We first use the \emph{PSF} task to extract from the $R$ and $Z$ mosaics the relative PSF models. 
Then we inserted in each mosaic 1000 artificial sources using the \emph{addstar} task; 
this number should not alter the crowding statistics in the images significantly. 
The profile for the artificial sources was generated by using the relative PSF model, while 
their positions are randomly distributed over the entire area of the mosaic and their 
magnitudes range uniformly between the detection and the saturation limits in the relative filter 
(see Table~\ref{tab:mag_limits}). 

\begin{figure} 
\centering
\includegraphics[width=9cm,height=7cm]{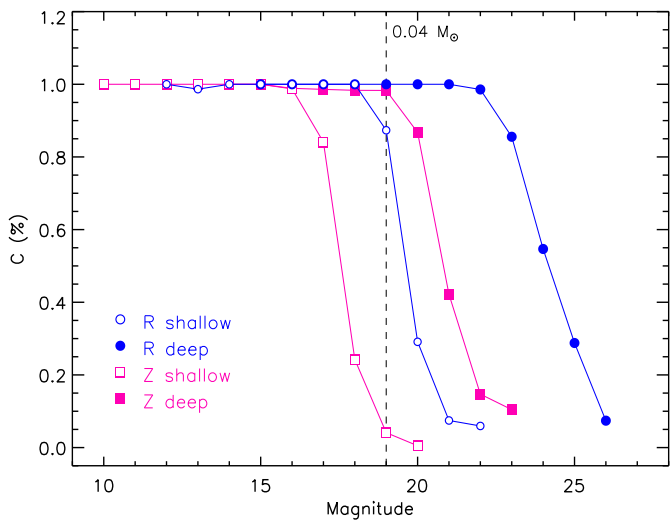}
\caption{Completeness (C) plot for extraction of artificial stars from our 
``deep'' and ``shallow'' mosaics for the $R$ and $Z$ bands.}
\label{fig_compl}
\end{figure}

Figure~\ref{fig_compl} shows the fraction of recovered artificial objects as a function 
of magnitude from both the deep and shallow mosaics for the $R$ and $Z$ filters. 
The corresponding magnitude limits at 95\% 
completeness level (C=95\%) are reported in Table~\ref{tab:mag_limits}. 

%---------------------------------- completeness-------------------
%-------------------------------------------------------------- 

In the absence of extinction (A$_V$=0), assuming a typical age for Serpens members of 2-6~Myr \citep{Oli09}, 
a distance of 260~pc \citep{Str96} and using the theoretical isochrones and evolutionary tracks by \citet{Bar98} and 
\citet{Cha00} {\bf and the completeness limits from Table~\ref{tab:mag_limits}}, 
our optical survey would be complete down to 0.04~$M_{\odot}$ at the 95\% level, 
{\bf i.e. well below the Hydrogen burning limit ($\sim$0.08~M$_\odot$)}.
However, as shown by \citet{Har07} in their Fig.~2, the visual extinction toward the area observed 
in Serpens has a typical value of A$_V \approx$7~mag. 
Adopting this extinction, we estimate a 95\% completeness level down to $M\approx$0.1~M$_\odot$,
 
The full catalog is downloadable from {\bf \emph{http://cdsarc.u-strasbg.fr/cgi-bin/VizieR}}.

\section{On the disk-less population in Serpens \label{use}}

The observations presented in this paper are part of the c2d complementary work in Serpens. 
The optical/NIR catalog described in Sect.~\ref{catalog} has been merged with the existing 
X-ray to millimeter wavelengths observations collected by the c2d Team for Serpens 
\citep{Goo04,Har07,Eno07,Oli09,Oli10}, in order to construct complete spectral 
energy distributions (SEDs) of the YSOs in the observed fields.
This dataset allows to anchor disk properties of the Serpens YSO sample to the 
mass, age, and evolutionary status of the central object, 
which is a critical point in the studies of envelopes/disks formation and 
evolution \citep[see, e.g.,][]{Kun06,Mey09,Pas09}. 
A number of papers based on the use of the c2d extensive dataset for Serpens
and presenting results on disk evolution are in preparation. 

As mentioned in Sect.~\ref{intro}, \citet{Com09} have recently identified a 
large population of young members of the Lupus dark cloud complex which seems 
to have lost their inner disks on a timescale of a few Myr or less. 
This discovery poses the important question of whether the existence of such disk-less 
young stars is the outcome of specific star-forming conditions in Lupus or 
similar populations exist in other regions.

\begin{figure} 
\centering
\includegraphics[width=8cm,height=7cm]{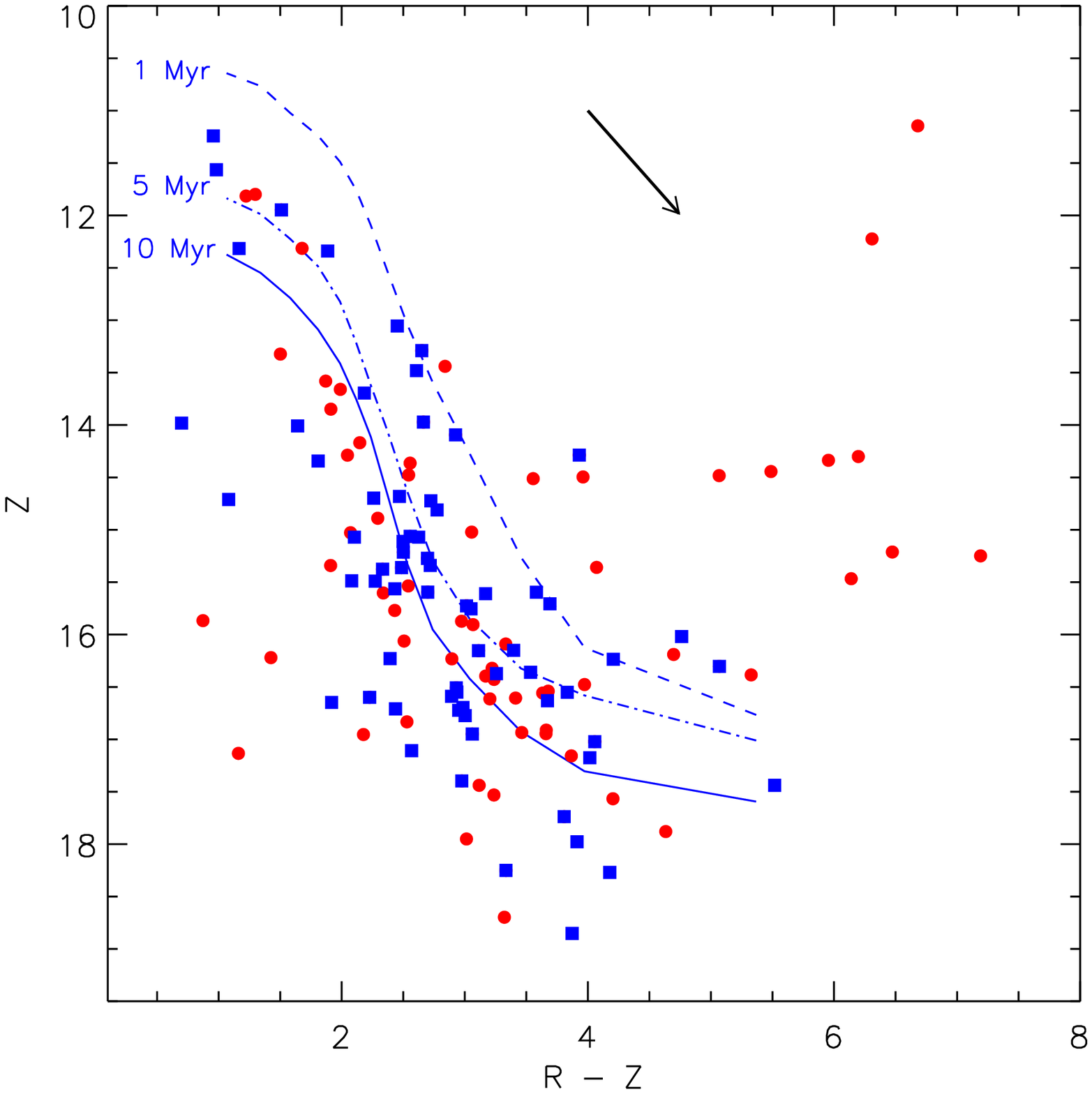}
\includegraphics[width=8cm,height=7cm]{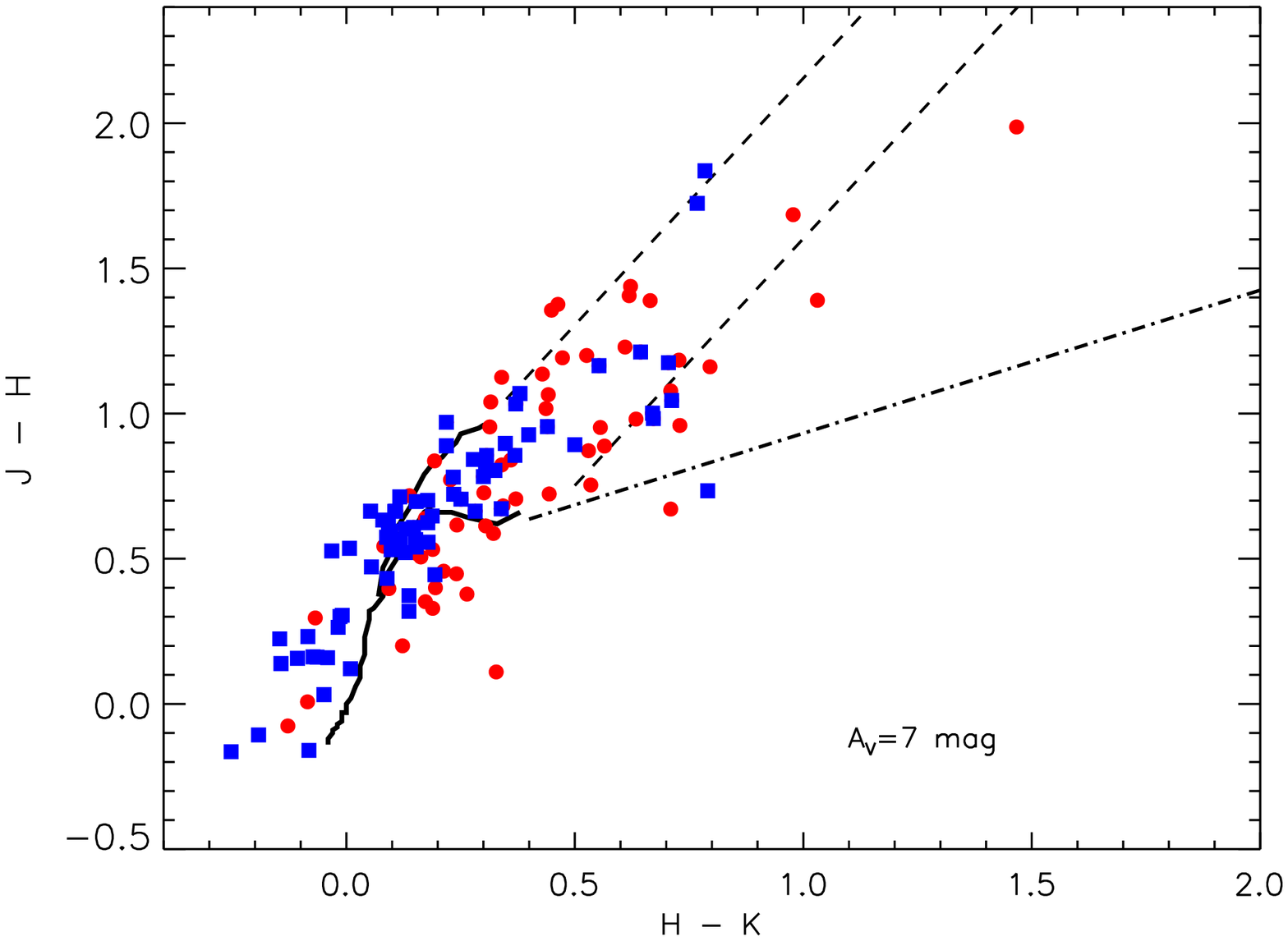}
\caption{$Z$ vs. $R-Z$ and $J-H$ vs. $H-K$ diagrams for the YSOs in Serpens 
selected by the c2d survey (circles) and the optical/NIR counterparts of the X-ray sources identified by 
\citet{Bro10} (squares). {\bf Upper panel:} The lines are the PMS isochrones 
by \citet{Bar98} and \citet{Cha00} transformed by us into the SDSS photometric system. The 
arrow represents {\bf A$_V$=2 reddening vector} \citep{Wei01}. 
{\bf Lower panel:} The solid curve shows the relation between the 
colour indices for main sequence stars (lower branch) and giants (upper branch), 
together with the relative reddening bands (dashed lines). 
{\bf The dash-dot line} is the T~Tauri star locus by \citet{Mey97}. 
The object magnitudes are derreddened using the typical extinction 
in the observed area (A$_V \approx$7~mag) and the extinction curve by \citet{Wei01}.}
\label{CMD_CC}
\end{figure}

XMM-Newton data obtained in April 2007 and April 2008 of the same field 
observed by the c2d survey in Serpens revealed a sample of new sources, 
half of which have no counterpart in the c2d catalog and are 
mostly candidate WTTSs \citep{Bro10}. 
For $\sim$44\% of the objects in this sample, 
we find optical and NIR colours consistent with those of young 
objects with no prominent IR excess. Indeed, they occupy in the $Z$ vs. $R-Z$ 
(Fig.~\ref{CMD_CC}, upper panel) the same locus as the YSOs identified by the c2d survey 
\citep{Har07}, which corresponds to an age between 1 and 10 Myr according to pre-main sequence (PMS) 
isochrones by \citet{Bar98} and \citet{Cha00} {\bf and is consistent with the ages 
of the c2d YSOs found by \citet{Oli09}}. The isochrones have 
been matched to the SDSS photometric system following the procedure described in 
Appendix~B by \citet{Spe07}. On the $J-H$ vs. $H-K$ diagram the c2d YSOs follow, as expected, 
the dwarf stars locus with many of them presenting IR excess with respect to this locus; 
the X-ray sources mainly concentrate on the field dwarfs locus, as expected 
for objects with no disk. 
For the remaining objects in the X-ray source sample we could not perform this analysis 
because they are either saturated or not detected in our optical/NIR survey, while a few of them 
are out of the observed field.

%---------------------------------- CMD and CC------------------

%-------------------------------------------------------------- 

To further investigate the disk-less population in Serpens, we tried to apply 
to our dataset the novel $S$-parameter method by \citet{Com09}, 
which allows the identification of possible young stars and substellar objects 
based on their optical/NIR photospheric fluxes, 
independently of the display of signposts of youth, 
such as IR excess emission or strong H$\alpha$ emission. 
The $S$-parameter method simultaneously estimates the visual extinction ($A_V$), 
effective temperature ($T_{\rm eff}$) and a wavelength-independent scaling factor ($S$), 
containing the dependency on the actual distance and radius of the star, 
by fitting a grid of stellar photosphere models to the observed SED. 
Using the Galactic star counts model by \citet{Wai92}, 
\citet{Com09} demonstrated that members of nearby young star forming regions 
at a given temperature ($<$4000~K) and 
within a restricted set of ages ($<$20~Myr) are characterised by values of 
$S$ virtually unreachable by non-members of similar temperature, 
{\bf both foreground dwarfs and background giants}. 
This method is expected to work reasonably well for the detection of cool populations 
associated to star-forming regions located in the $\sim$100-300~pc 
distance range from the Sun \citep[see Sect.~5 by][]{Com09} and, indeed, 
it has been successfully applied to a number of nearby star forming regions: 
Lupus \citep{Com09}, Cha~I (Lop\`ez Mart\`i, private communication), Cr~A \citep{Lop10} 
and Cha~II, where it provides the same results as in \citet{Spe07}. 
We applied the $S$-parameter method to the optical/NIR dataset for Serpens; 
we find that in this case the $S$-parameter histogram at any temperature range shows no clear 
separation between the member candidates and the background/foreground contaminants and,
as a consequence, the method produces an unreliable sample of member candidates, 
{\bf which is also not compatible with the XMM-detected sources in the same area of the sky}. 
There are recent indications \citep[see Sect.~2.3 by][]{Mer08} that Galactic models 
may fail at reproducing stellar counts at low galactic latitude and, indeed, 
Serpens lies very close to the galactic plane 
($b \approx$+5.4~deg). Moreover, Serpens is located at 260~pc from the Sun, 
i.e. at the limit of the validity range of the $S$-parameter method.
Since all the other clouds on which the method has been tested are closer to the Sun 
and have higher galactic latitude than Serpens, 
we conclude that the unlucky combination of distance and position of this cloud 
prevents the $S$-parameter method from giving useful results.

\section{Conclusions \label{conclu}}

We presented an optical/NIR catalog ($R$ and $Z$ filters) 
of 26524 point-like sources in 0.96~square degrees in the 
direction of Serpens down to R$\approx$25. 
These data were collected using the WFC camera at the INT 
within the frame of the Spitzer c2d survey in Serpens. 
The catalog was also complemented with $JHK_S$ photometry from 2MASS and 
has been merged with the existing X-ray to millimeter wavelengths 
observations collected by the c2d Team for Serpens 
to study the envelop/disk formation and evolution 
and its dependency on the stellar properties. 
A number of paper based on this comprehensive catalog are now published or in preparation 
\citep{Mer10a,Mer10b,Oli10,Bro10}.

In this paper we used the optical/NIR catalog to investigate the 
disk-less population in Serpens. 
Because of the distance and low galactic latitude of this cloud, 
the $S$-parameters method by \citet{Com09}, suitable for the identification of young objects 
independently of any signature of youth, fails in distinguishing young cloud members 
from field contaminants.
However, a sample of new WTTS candidates has been identified in Serpens on the basis of 
XMM-Newton observations. Our optical/NIR photometry suggests a very young age ($\lesssim$10~Myr) 
and no NIR excess emission for about 44\% of them, supporting their WTTS nature.

\begin{acknowledgements}

This publication makes use of data products from the Two Micron All Sky
Survey, which is a joint project of the University of Massachusetts and
the Infrared Processing and Analysis Center/California Institute of
Technology, funded by NASA and the National Science Foundation. 
We also acknowledge extensive use of the SIMBAD database, 
operated at CDS Strasbourg. 
We thank Ignas Snellen and the students from Leiden Observatory for taking 
these INT data in service time. 
We thank Mario Radowich for the concession of ASTROMETRIX. We are also grateful to 
F. Comer\'on, J.M. Alcal\'a, H. Bouy, B. Lop\'ez Mart\'i, R. Jayawardhana and the Spitzer c2d Team 
for useful discussions and suggestions.
We are also grateful to many others, in particular to Salvatore Spezzi.

\end{acknowledgements}

\bibliographystyle{aa}

\end{document}